\documentstyle[twoside,graphicx]{article}

\catcode`\@=11
\long\def\@makefntext#1{
\protect\noindent \hbox to 3.2pt {\hskip-.9pt  
$^{{\eightrm\@thefnmark}}$\hfil}#1\hfill}		

\def\thefootnote{\fnsymbol{footnote}}
\def\@makefnmark{\hbox to 0pt{$^{\@thefnmark}$\hss}}	
	
\def\ps@myheadings{\let\@mkboth\@gobbletwo
\def\@oddhead{\hbox{}
\rightmark\hfil\eightrm\thepage}   
\def\@oddfoot{}\def\@evenhead{\eightrm\thepage\hfil
\leftmark\hbox{}}\def\@evenfoot{}
\def\sectionmark##1{}\def\subsectionmark##1{}}

\oddsidemargin=\evensidemargin
\renewcommand{\thefootnote}{\fnsymbol{footnote}}

\newcounter{sectionc}\newcounter{subsectionc}\newcounter{subsubsectionc}
\renewcommand{\section}[1] {\vspace{12pt}\addtocounter{sectionc}{1} 
\setcounter{subsectionc}{0}\setcounter{subsubsectionc}{0}\noindent 
	{\tenbf\thesectionc. #1}\par\vspace{5pt}}
\renewcommand{\subsection}[1] {\vspace{12pt}\addtocounter{subsectionc}{1} 
	\setcounter{subsubsectionc}{0}\noindent 
	{\bf\thesectionc.\thesubsectionc. {\kern1pt \bfit #1}}\par\vspace{5pt}}
\renewcommand{\subsubsection}[1] {\vspace{12pt}\addtocounter{subsubsectionc}{1}
	\noindent{\tenrm\thesectionc.\thesubsectionc.\thesubsubsectionc.
	{\kern1pt \tenit #1}}\par\vspace{5pt}}
\newcommand{\nonumsection}[1] {\vspace{12pt}\noindent{\tenbf #1}
	\par\vspace{5pt}}

\newcounter{appendixc}
\newcounter{subappendixc}[appendixc]
\newcounter{subsubappendixc}[subappendixc]
\renewcommand{\thesubappendixc}{\Alph{appendixc}.\arabic{subappendixc}}
\renewcommand{\thesubsubappendixc}
	{\Alph{appendixc}.\arabic{subappendixc}.\arabic{subsubappendixc}}

\renewcommand{\appendix}[1] {\vspace{12pt}
        \refstepcounter{appendixc}
        \setcounter{figure}{0}
        \setcounter{table}{0}
        \setcounter{lemma}{0}
        \setcounter{theorem}{0}
        \setcounter{corollary}{0}
        \setcounter{definition}{0}
        \setcounter{equation}{0}
        \renewcommand{\thefigure}{\Alph{appendixc}.\arabic{figure}}
        \renewcommand{\thetable}{\Alph{appendixc}.\arabic{table}}
        \renewcommand{\theappendixc}{\Alph{appendixc}}
        \renewcommand{\thelemma}{\Alph{appendixc}.\arabic{lemma}}
        \renewcommand{\thetheorem}{\Alph{appendixc}.\arabic{theorem}}
        \renewcommand{\thedefinition}{\Alph{appendixc}.\arabic{definition}}
        \renewcommand{\thecorollary}{\Alph{appendixc}.\arabic{corollary}}
        \renewcommand{\theequation}{\Alph{appendixc}.\arabic{equation}}
        \noindent{\tenbf Appendix \theappendixc #1}\par\vspace{5pt}}
\newcommand{\subappendix}[1] {\vspace{12pt}
        \refstepcounter{subappendixc}
        \noindent{\bf Appendix \thesubappendixc. {\kern1pt \bfit #1}}
	\par\vspace{5pt}}
\newcommand{\subsubappendix}[1] {\vspace{12pt}
        \refstepcounter{subsubappendixc}
        \noindent{\rm Appendix \thesubsubappendixc. {\kern1pt \tenit #1}}
	\par\vspace{5pt}}

\newcommand{\textlineskip}{\baselineskip=13pt}
\newcommand{\smalllineskip}{\baselineskip=10pt}

\def\eightcirc{
\begin{picture}(0,0)
\put(4.4,1.8){\circle{6.5}}
\end{picture}}
\def\eightcopyright{\eightcirc\kern2.7pt\hbox{\eightrm c}} 

\newcommand{\copyrightheading}[1]
	{\vspace*{-2.5cm}\smalllineskip{\flushleft
	{\footnotesize International Journal of Modern Physics A, #1}\\
	{\footnotesize $\eightcopyright$\, World Scientific Publishing
	 Company}\\
	 }}

\def\abstracts#1#2#3{{
	\centering{\begin{minipage}{4.5in}\baselineskip=10pt\footnotesize
	\parindent=0pt #1\par 
	\parindent=15pt #2\par
	\parindent=15pt #3
	\end{minipage}}\par}} 

\renewenvironment{thebibliography}[1]
	{\frenchspacing
	 \ninerm\baselineskip=11pt
	 \begin{list}{\arabic{enumi}.}
	{\usecounter{enumi}\setlength{\parsep}{0pt}
	 \setlength{\leftmargin 12.7pt}{\rightmargin 0pt} 
	 \setlength{\itemsep}{0pt} \settowidth
	{\labelwidth}{#1.}\sloppy}}{\end{list}}

\newcounter{itemlistc}
\newcounter{romanlistc}
\newcounter{alphlistc}
\newcounter{arabiclistc}

\newcommand{\fcaption}[1]{
        \refstepcounter{figure}
        \setbox\@tempboxa = \hbox{\footnotesize Fig.~\thefigure. #1}
        \ifdim \wd\@tempboxa > 5in
           {\begin{center}
        \parbox{5in}{\footnotesize\smalllineskip Fig.~\thefigure. #1}
            \end{center}}
        \else
             {\begin{center}
             {\footnotesize Fig.~\thefigure. #1}
              \end{center}}
        \fi}

\newcommand{\tcaption}[1]{
        \refstepcounter{table}
        \setbox\@tempboxa = \hbox{\footnotesize Table~\thetable. #1}
        \ifdim \wd\@tempboxa > 5in
           {\begin{center}
        \parbox{5in}{\footnotesize\smalllineskip Table~\thetable. #1}
            \end{center}}
        \else
             {\begin{center}
             {\footnotesize Table~\thetable. #1}
              \end{center}}
        \fi}

\def\@citex[#1]#2{\if@filesw\immediate\write\@auxout
	{\string\citation{#2}}\fi
\def\@citea{}\@cite{\@for\@citeb:=#2\do
	{\@citea\def\@citea{,}\@ifundefined
	{b@\@citeb}{{\bf ?}\@warning
	{Citation `\@citeb' on page \thepage \space undefined}}
	{\csname b@\@citeb\endcsname}}}{#1}}

\newif\if@cghi
\def\cite{\@cghitrue\@ifnextchar [{\@tempswatrue
	\@citex}{\@tempswafalse\@citex[]}}
\def\citelow{\@cghifalse\@ifnextchar [{\@tempswatrue
	\@citex}{\@tempswafalse\@citex[]}}
\def\@cite#1#2{{$\null^{#1}$\if@tempswa\typeout
	{IJCGA warning: optional citation argument 
	ignored: `#2'} \fi}}

\def\pmb#1{\setbox0=\hbox{#1}
	\kern-.025em\copy0\kern-\wd0
	\kern.05em\copy0\kern-\wd0
	\kern-.025em\raise.0433em\box0}

\def\fnt#1#2{\footnotetext{\kern-.3em
	{$^{\mbox{\scriptsize #1}}$}{#2}}}

\def\fpage#1{\begingroup
\voffset=.3in
\thispagestyle{empty}\begin{table}[b]\centerline{\footnotesize #1}
	\end{table}\endgroup}

\def\runninghead#1#2{\pagestyle{myheadings}
\markboth{{\protect\footnotesize\it{\quad #1}}\hfill}
{\hfill{\protect\footnotesize\it{#2\quad}}}}
\font\tenrm=cmr10
\font\tenit=cmti10 
\font\tenbf=cmbx10
\font\bfit=cmbxti10 at 10pt
\font\ninerm=cmr9

\font\eightrm=cmr8

\def\qed{\hbox{${\vcenter{\vbox{			
   \hrule height 0.4pt\hbox{\vrule width 0.4pt height 6pt
   \kern5pt\vrule width 0.4pt}\hrule height 0.4pt}}}$}}

\renewcommand{\thefootnote}{\fnsymbol{footnote}}	

\begin{document}

\runninghead{QCD measurements in photon-photon collisions at LEP}{QCD 
measurements in photon-photon collisions at LEP}

\normalsize\textlineskip
\thispagestyle{empty}
\setcounter{page}{1}

\copyrightheading{}			

\vspace*{0.7truein}

\fpage{1}
\centerline{\bf QCD MEASUREMENTS IN PHOTON-PHOTON COLLISIONS AT LEP
\footnote{Talk presented at DPF 2000, Columbus, Ohio, August 2000.}
}
\vspace*{0.37truein}
\centerline{\footnotesize \'AKOS CSILLING\footnote{The author thanks the UK
Particle Physics and Astronomy Research Council for their support.\break
Permanent address: KFKI Research Institute for Particle and Nuclear Physics,
Budapest, P.O.Box~49, H-1525, Hungary; supported by the Hungarian Foundation
for Scientific Research, OTKA~F-023259.}}
\vspace*{0.015truein}
\centerline{\footnotesize\it  Department of Physics and Astronomy, 
UCL,}
\baselineskip=10pt
\centerline{\footnotesize\it  Gower Street, London WC1E 6BT, United Kingdom
}

\vspace*{0.21truein}
\abstracts{
An overview of the latest results of the LEP collaborations on QCD measurements 
in photon-photon collisions
is presented, including measurements of the total hadronic cross-section, the
production of heavy quarks and dijets and the structure functions of 
real and virtual photons.
}{}{}


\vspace*{1pt}\textlineskip	
\section{Introduction}	
\vspace*{-0.5pt}

\noindent Hadronic photon-photon interactions are responsible for the vast
majority of hadron production in $\mathrm{e^+e^-}$ collisions at  LEP2
energies.  While soft processes have a large contribution to the total
cross-section and non-perturbative  phenomenological models are invoked to
describe this data,  various other QCD processes can also be measured in the
perturbative region and compared to NLO predictions.

The high energy and large luminosity of LEP provides very high statistics
samples of photon-photon interactions, allowing the improvement of earlier
measurements both in terms of statistical power and in terms of the covered
phase-space, as well as the measurement of hitherto unmeasured rare processes.
The continuous improvement of the available Monte Carlo
models and better data analysis techniques contribute to the reduction of
systematic uncertainties.

An overview of the latest results of the LEP collaborations in the study of QCD
processes in photon-photon interactions is presented.

\textheight=7.8truein
\setcounter{footnote}{0}
\renewcommand{\thefootnote}{\alph{footnote}}

\section{Total Hadronic Cross-section}

\begin{figure}
\includegraphics[width=0.49\textwidth,height=0.45\textwidth]{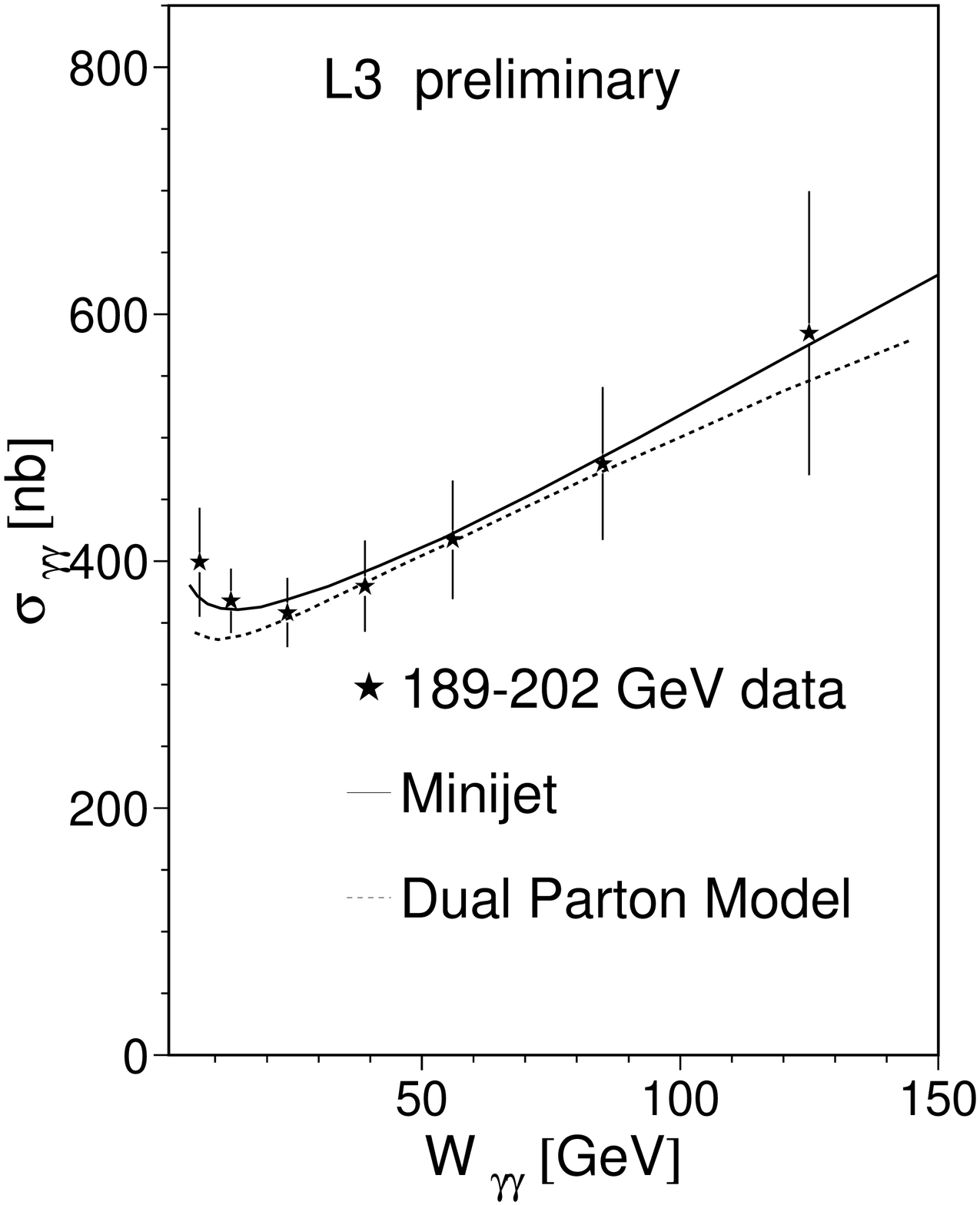}
\includegraphics[width=0.49\textwidth,height=0.45\textwidth]{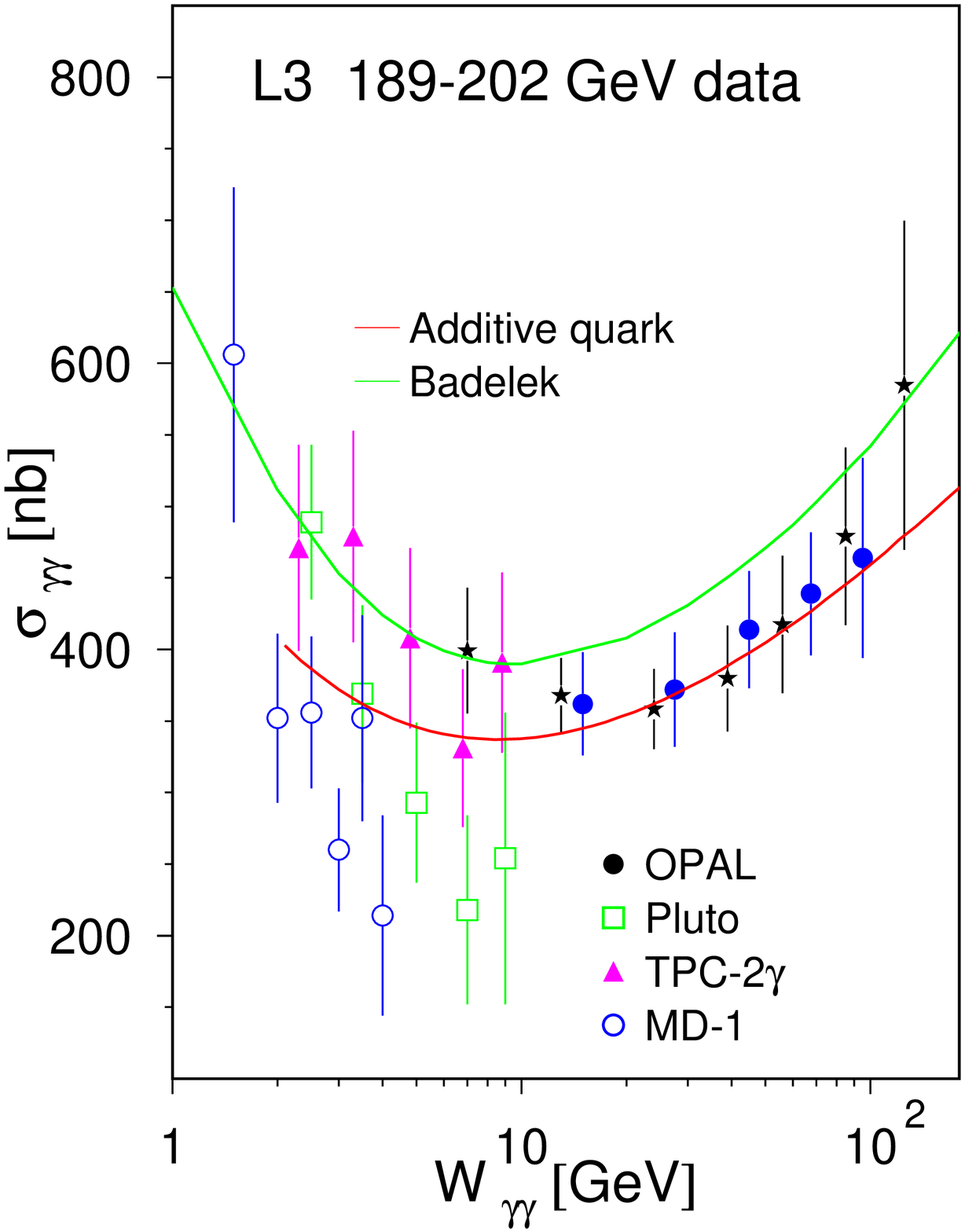}
\fcaption{Total cross-section of quasi-real photon-photon scattering measured by
L3, compared with other experiments and various QCD 
models.}
\label{fig:total}
\end{figure}

\noindent The  total hadronic cross-section of quasi-real photon-photon
collisions has been measured at LEP by L3\cite{L3total} and
OPAL\cite{opaltotal} up to  a photon-photon invariant mass of
$W_{\gamma\gamma}=\sqrt{s_{\gamma\gamma}}=145$
GeV, as shown in Figure~\ref{fig:total}. The visible invariant mass of each
event is calculated from the particles observed in the detector, while the true
$W_{\gamma\gamma}$ distribution has to be recovered by a statistical unfolding
procedure which introduces correlations between data points. The photon
flux is calculated by numerical integration over each bin, taking into account
the finite virtuality of the incoming photons.  Most of the systematic
uncertainties involved in the measurement are due to our limited understanding
of soft and diffractive processes, manifested in a large model-dependence of
the unfolding procedure. 

The total cross-section  increases as a function of
$W_{\gamma\gamma}$ faster than expected from the universal Donnachie-Landshoff
fit. The observed energy dependence can be reproduced by introducing a hard
Pomeron component to the fit, or by QCD models which attribute an important
contribution to the hard scattering of the partons inside the photon.

\section{Heavy Quark Production}

\noindent The measurement of heavy quark production in photon-photon collisions
provides a better test of QCD, because the large physical scale of the charm and
beauty quark masses makes  perturbative calculations more reliable.

\begin{figure}[bt!]
\null\vspace{-8mm}\null
\includegraphics[width=0.49\textwidth,height=0.45\textwidth]{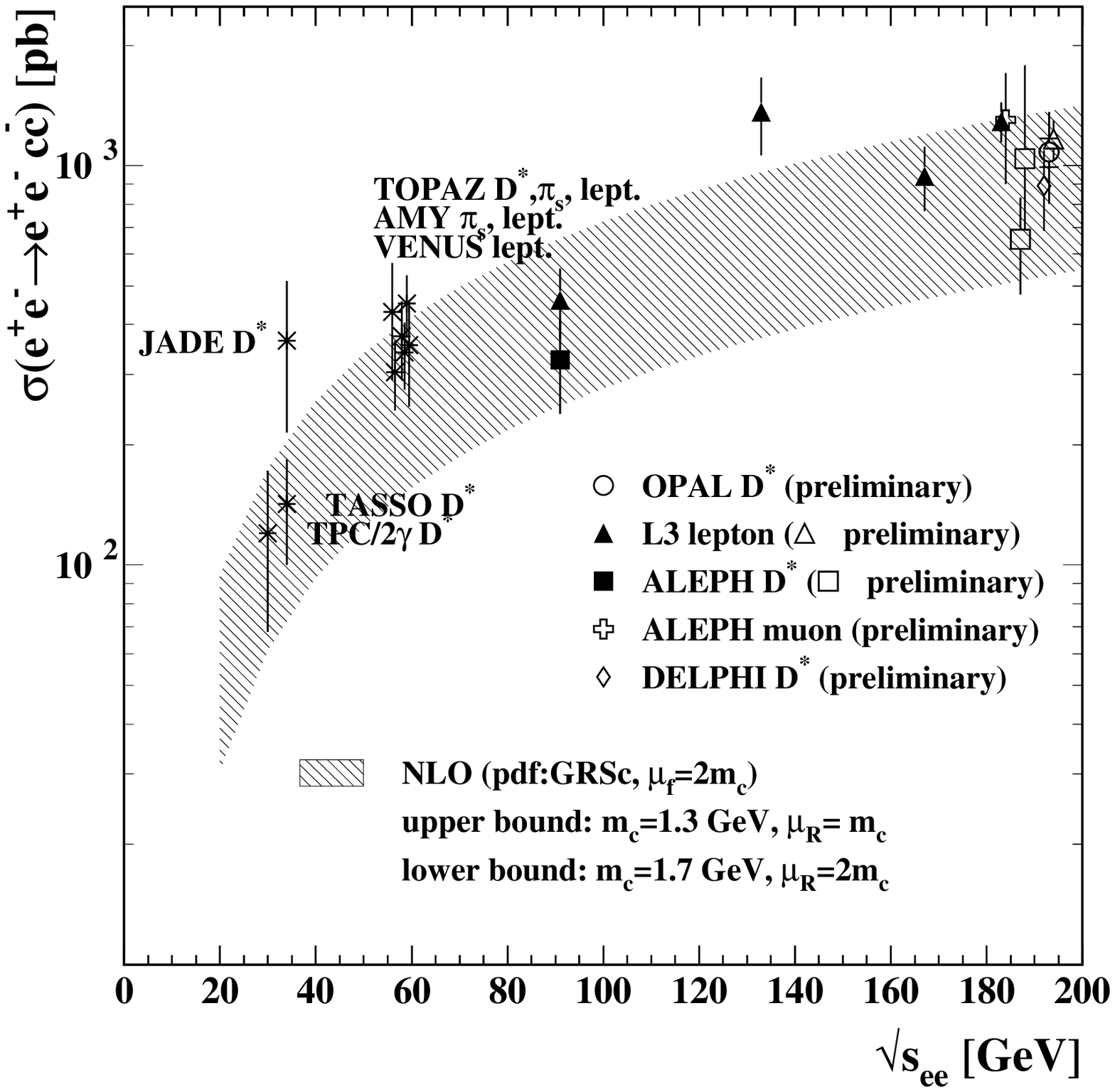}
\includegraphics[width=0.49\textwidth,height=0.48\textwidth]{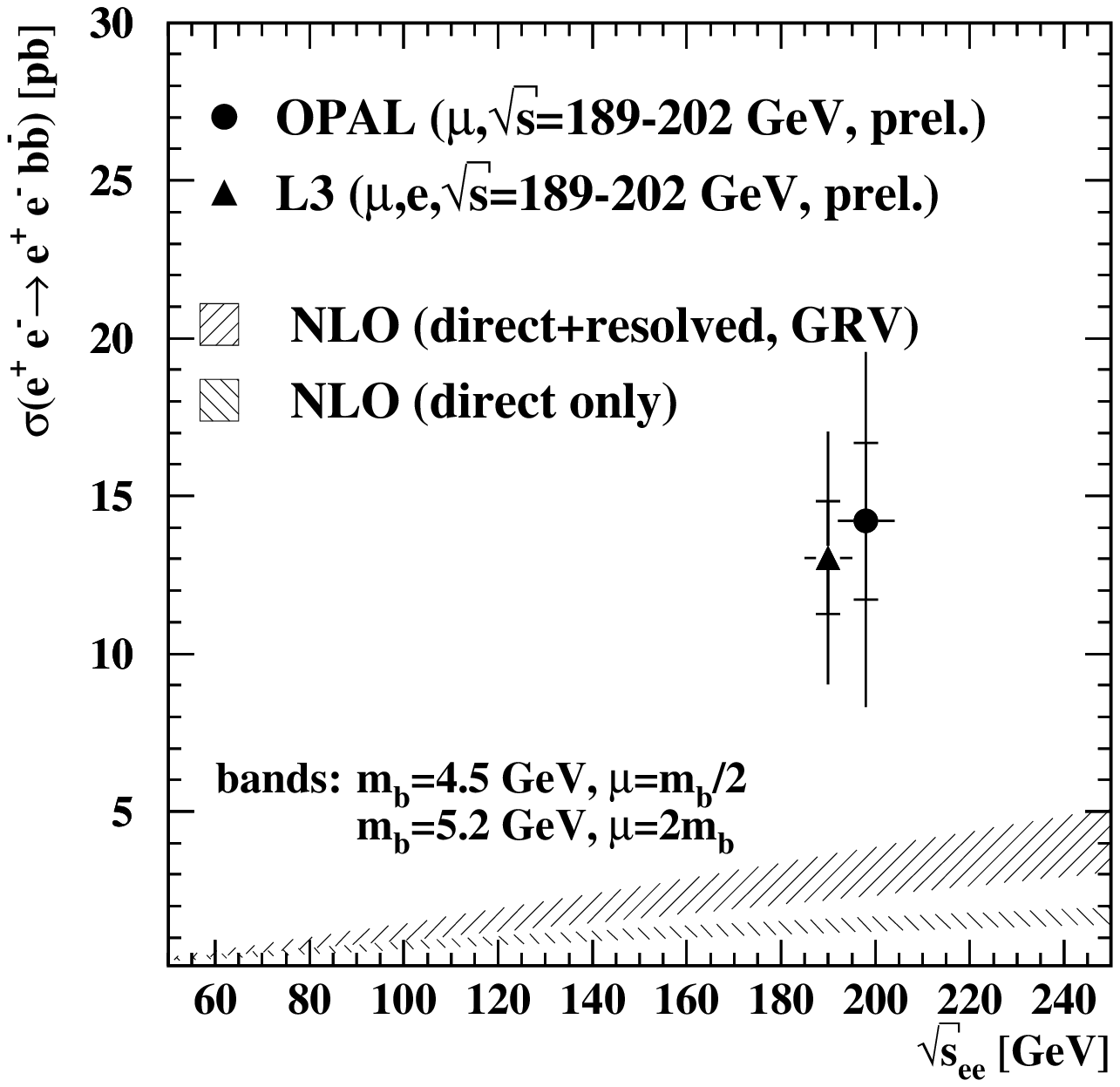}
\fcaption{Total cross-section of charm and beauty production in photon-photon
collisions.}
\label{fig:charm}
\label{fig:bottom}
\end{figure}

Charm production has been
measured\cite{alephheavy,delphiheavy,L3heavy,opalheavy} by the LEP
collaborations using both  the lepton-tag and the D$^*$-tag methods. The first
method uses the fact that most leptons observed in hadronic photon-photon
collisions originate from semileptonic charm decay. The second method is
based on the reconstruction of the $\mathrm{D^{*\pm}\to D^0\pi^\pm}$ decay,
which has a small combinatorial background due to the small difference of the
D$^{*\pm}$ and D$^0$ masses. Both methods give consistent results, and all the
recent measurements, shown in Figure~\ref{fig:charm} (left), lie within the
wide band of theoretical uncertainties. 

\begin{figure}[b!]
\includegraphics[width=0.49\textwidth,height=0.49\textwidth]{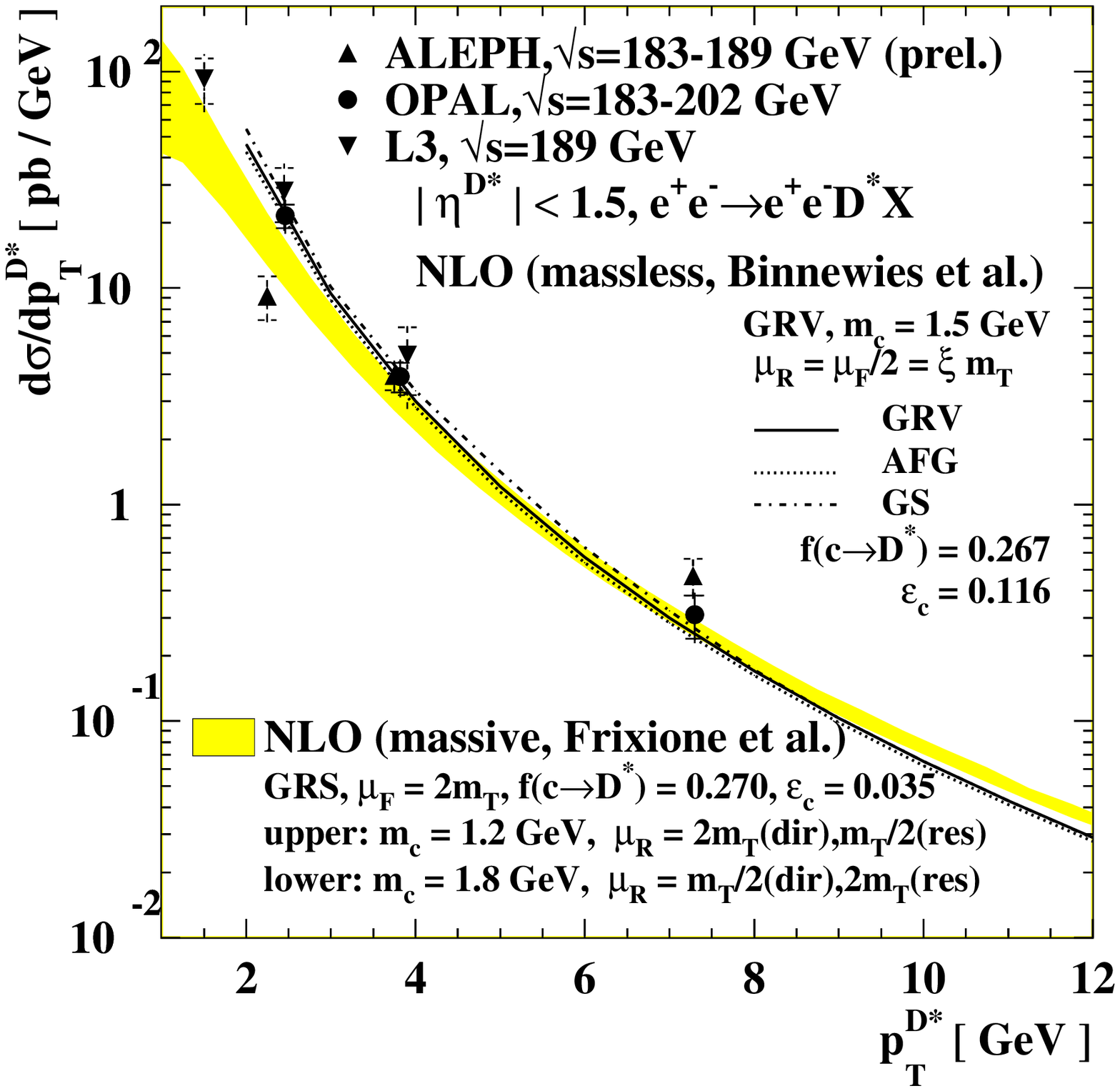}
\includegraphics[width=0.49\textwidth,height=0.49\textwidth]{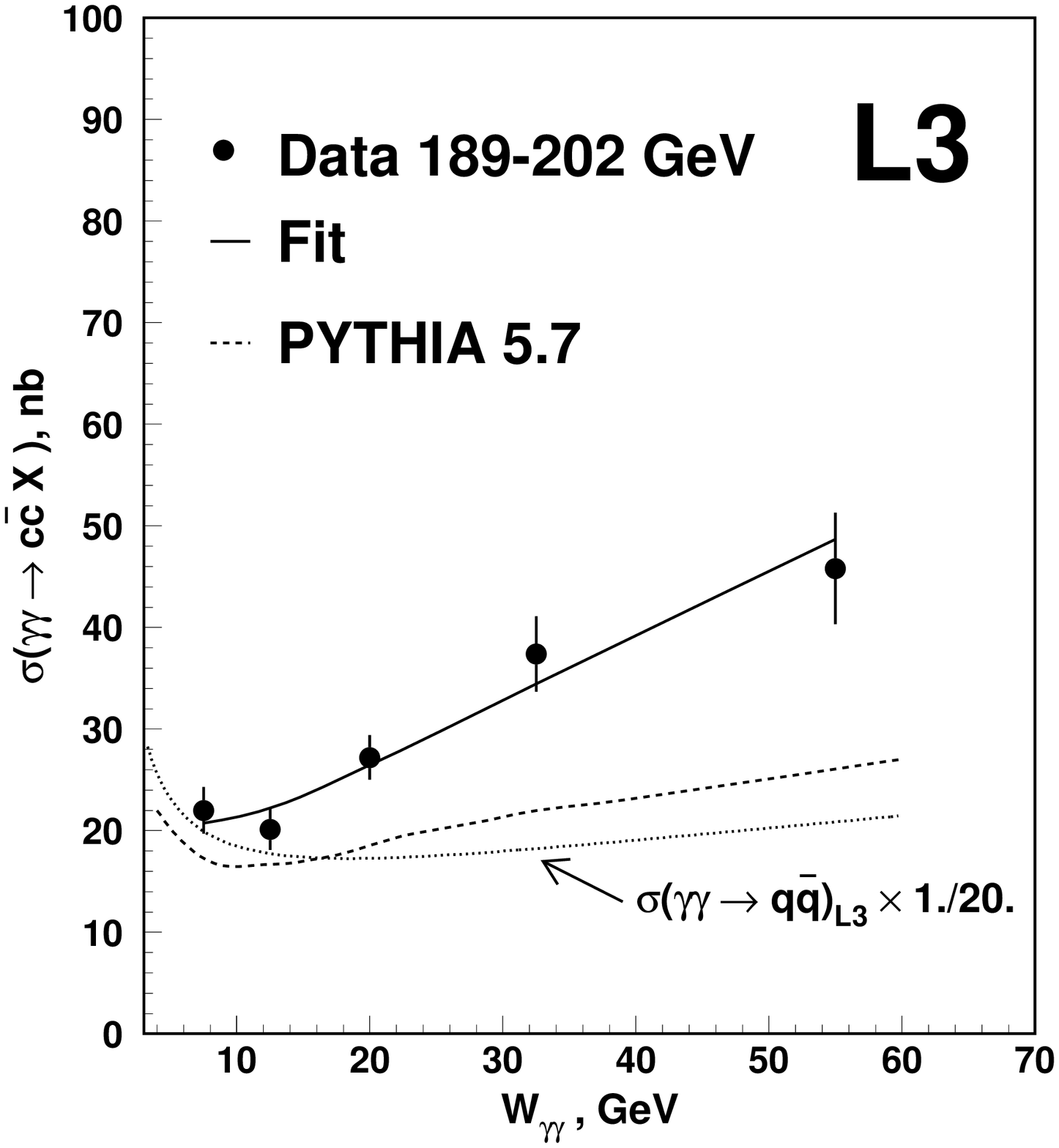}
\vspace*{2mm}
\fcaption{Differential cross-section of D$^*$  production in photon-photon
collisions, measured by ALEPH, L3 and OPAL, together with two different 
NLO calculations  (left) and the differential cross-section of charm
production as a function of $W_{\gamma\gamma}$ measured by L3 using the 
lepton-tag method.}
\label{fig:dstarpt}
\label{fig:ccw}
\end{figure}

One should note that the large errors both on the experimental measurements and
on the theoretical predictions can be reduced by restricting the considered
phase space, since the largest uncertainty originates from the extrapolation to
the kinematic region outside the detector acceptance. For example,  the
differential cross-section of D$^*$ production measured as a function of its
transverse momentum is shown in Figure~\ref{fig:dstarpt} (left).

The differential cross-section of charm production as a function of
$W_{\gamma\gamma}$ is shown in Figure~\ref{fig:ccw} (right). A fast rise is
observed towards high energies, much faster than in the total cross-section, in
agreement with measurements of the photoproduction of J$/\Psi$ mesons at HERA.

Beauty production is expected to be suppressed by more than two orders of
magnitude relative to charm production. It can be measured by exploiting the
harder transverse momentum spectrum of muons produced from semileptonic b
decays compared to other muon sources. The resulting cross-section of open
beauty production\cite{L3heavy,opalheavy} is  significantly underestimated by
the NLO prediction, while the L3 and OPAL measurements are in good agreement
with each other, as shown in Figure~\ref{fig:bottom} (right).

\section{Dijet Production}

\noindent The production of jets in photon-photon collisions  contains a hard
scale, the transverse momentum, $p_{\mathrm{T}}$, or transverse energy,
$E_{\mathrm{T}}$, of the jets, allowing the use of perturbative methods. Both
ALEPH\cite{alephdijet} and OPAL\cite{opaldijet} find that the NLO calculations
are able to describe the measured transverse energy and momentum distribution.

\begin{figure}[h]
\null\vspace{-7mm}\null
\includegraphics[width=0.49\textwidth,height=.51\textwidth]{pn443_15.eps}
\includegraphics[width=0.49\textwidth,height=.53\textwidth]{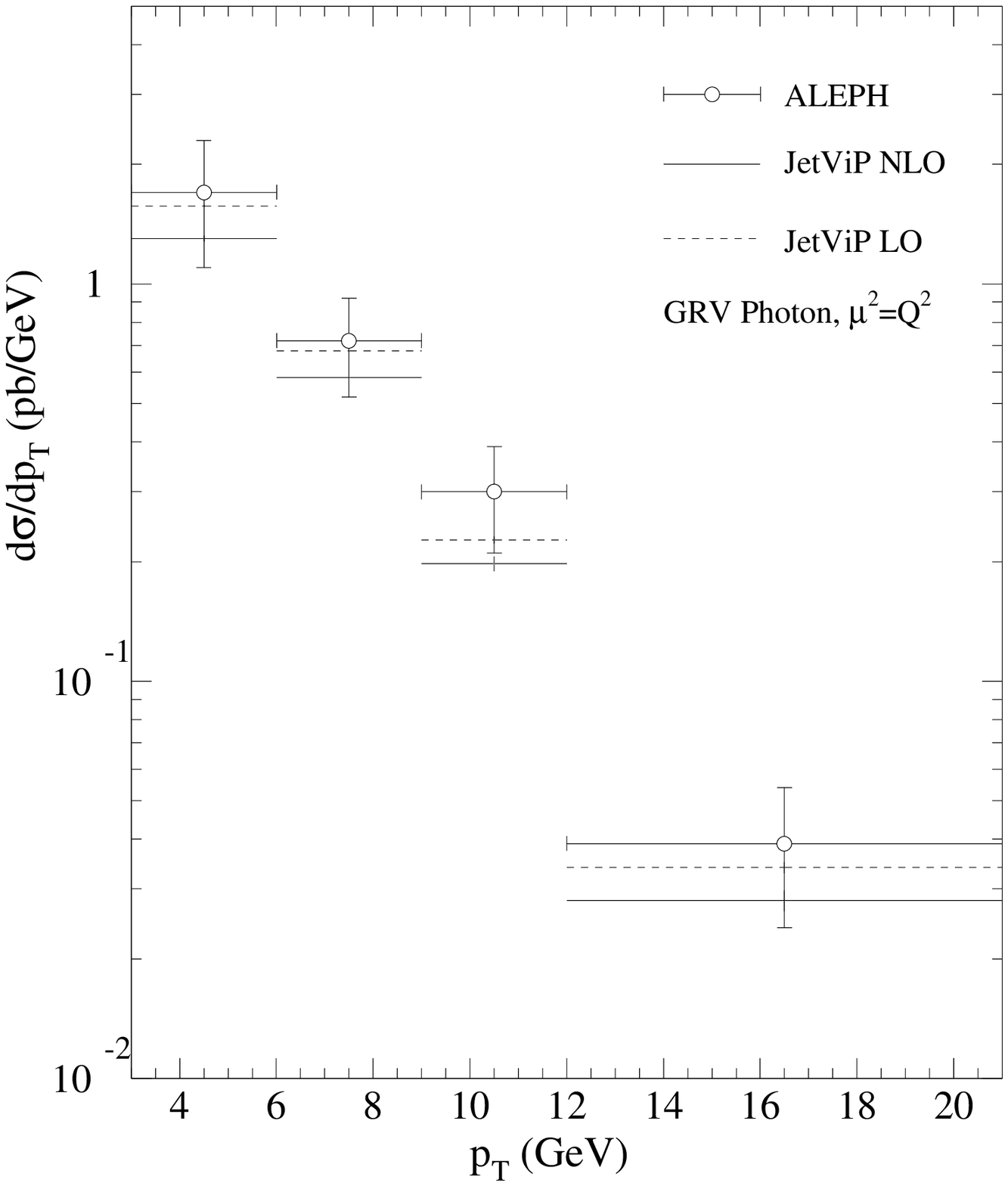}
\fcaption{Untagged dijet cross-sections as a function of $x_\gamma$ from  OPAL
(left) and tagged dijet cross-section as a function of $p_{\mathrm{T}}$  from
ALEPH (right) compared with NLO calculations.}\vspace*{-3mm}
\label{fig:dijet}
\label{fig:taggedjets}
\end{figure}

The quark and gluon contents of the photon  can be separated by determining
the fraction of the photon momenta carried by the jets, $x_\gamma$. Comparisons to the OPAL
data\cite{opaldijet} presented in Figure~\ref{fig:dijet} (left) show that at low $x_{\gamma}$ 
the gluon content of the photon is underestimated by the NLO calculation.
The differences observed at large $x_\gamma$ are due to hadronization effects
not taken into account in the  calculation.

Requiring a tagged photon in dijet events introduces another hard scale, the
virtuality of the probe photon, $Q^2$, which is typically of a similar order of
magnitude as $p_{\mathrm{T}}$. Despite the theoretically more complicated
nature of the problem, the ALEPH data\cite{alephdijettag} shown in
Figure~\ref{fig:taggedjets} (right) are well described by  NLO calculations.

\section{Single and Double-tag Measurements} 

\noindent  The structure of the photon can be measured in single-tag
$\gamma^*\gamma$  collisions, regarded as deep inelastic electron-photon
scattering. This classical measurement  was extended at LEP to lower $x$ and
larger $Q^2$ values than ever before. Requiring a second tagging electron
allows the study of the effective structure function of virtual photons if the
virtualities are very different, or the investigation of the dynamics of highly
virtual photon collisions if they are similar.

\begin{figure}[bt!]
\includegraphics[width=0.49\textwidth,height=.45\textwidth]{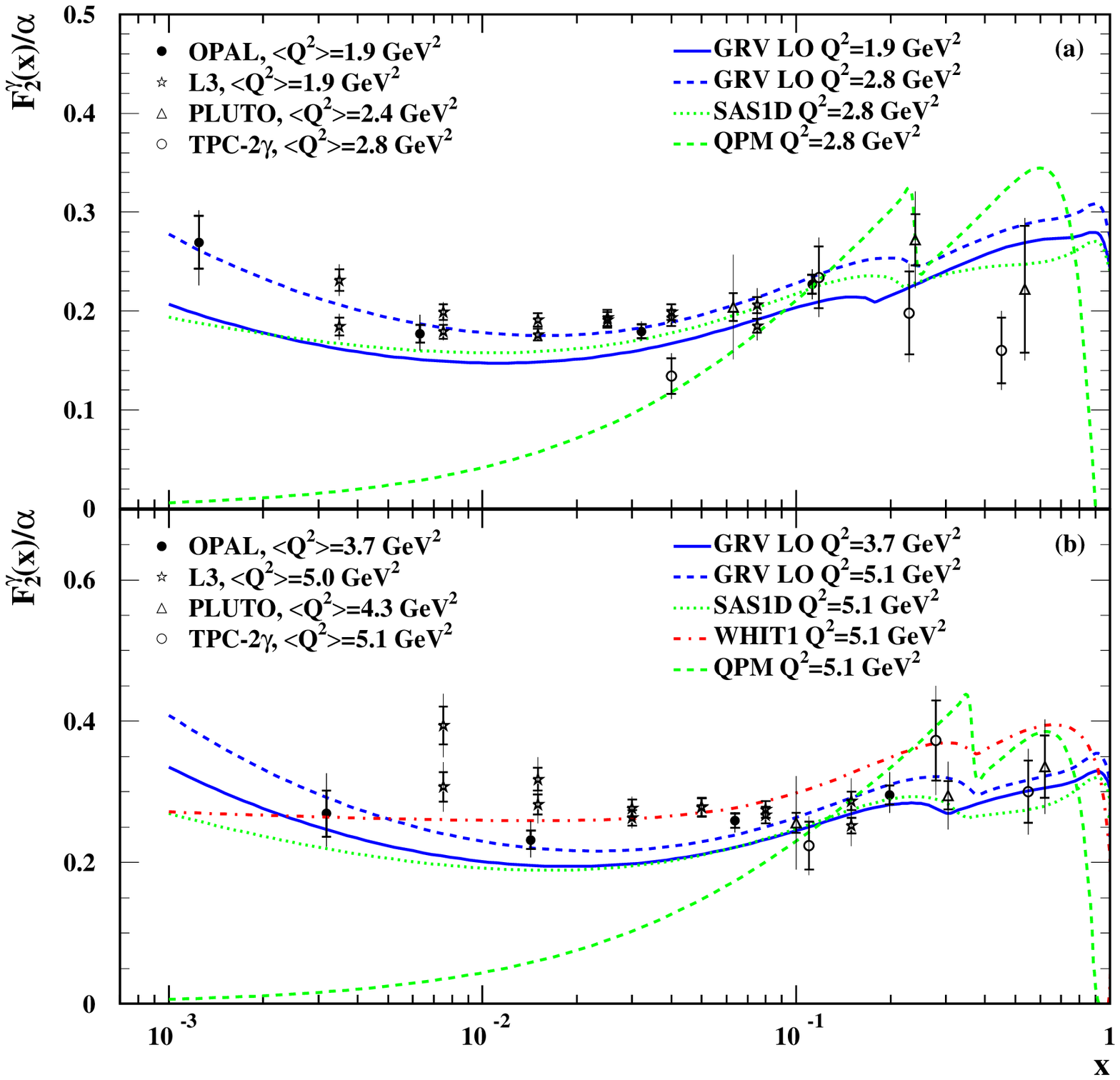}
\includegraphics[width=0.49\textwidth,height=.49\textwidth]{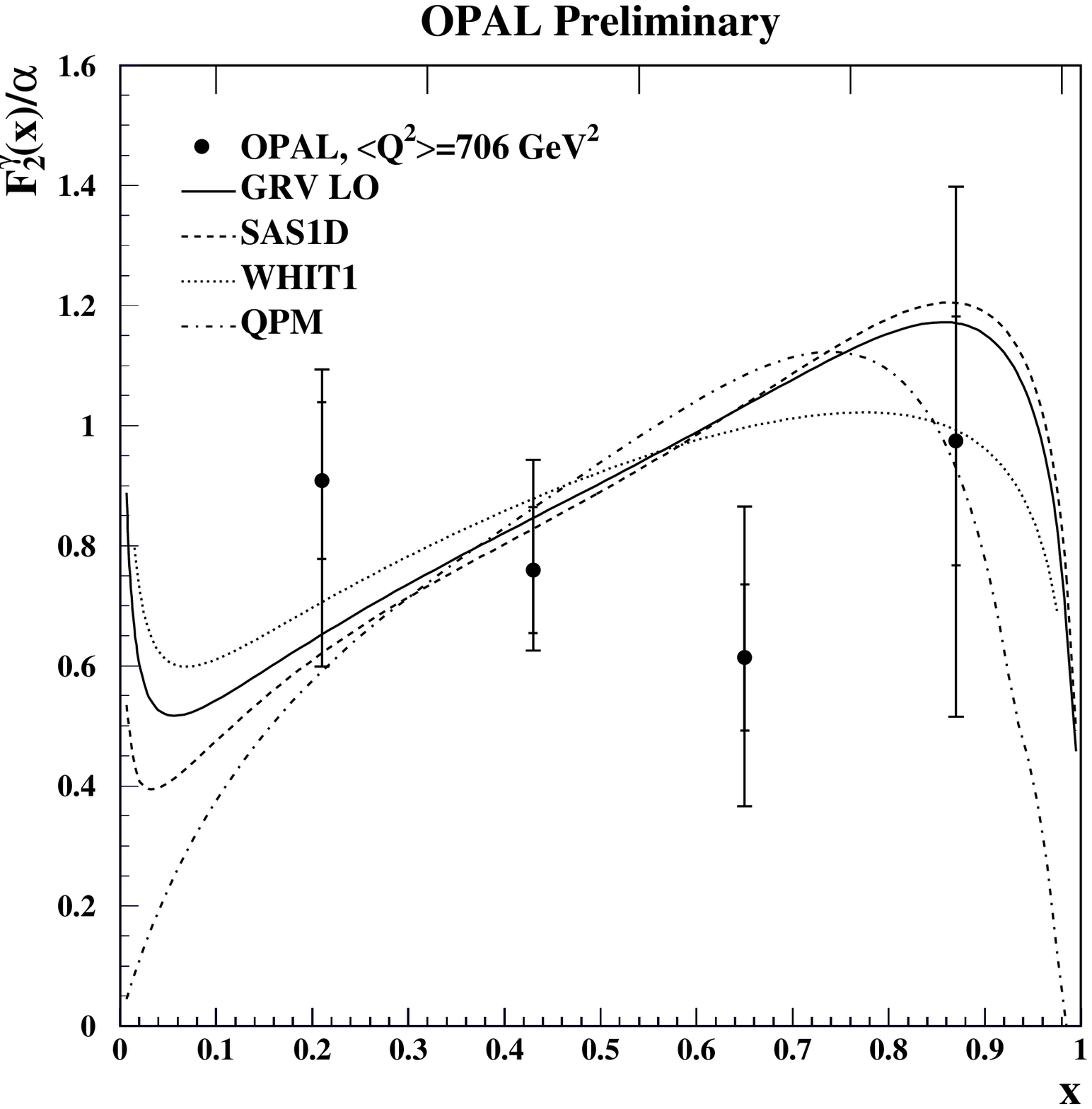}
\fcaption{The structure function of the photon at low $x$ (left) and high $Q^2$
(right) measured by OPAL.}
\label{fig:lowx}
\label{fig:highQ2}
\end{figure}

The recent measurement\cite{opallowx} of $F_2^{\gamma}$  at low-$x$  by OPAL
has benefited from significant improvements of the available  Monte Carlo
models and also from the use of new methods, such as multi-variable unfolding,
leading to smaller systematic uncertainties. While the data shown in
Figure~\ref{fig:lowx} (left) are completely consistent with the rise in
$F_2^\gamma$ towards lower values of $x$ expected from QCD calculations, the
obtained precision is still insufficient for a reliable confirmation. The high
$Q^2$ measurement\cite{L3highQ2} of L3, using LEP1 data, reaches
$Q^2=500$~GeV$^2$,  while OPAL reaches $Q^2=2200$~GeV$^2$ in a preliminary
analysis\cite{opalhighQ2} of the LEP2 data, shown in Figure~\ref{fig:highQ2}
(right). 

\begin{figure}[b!]
\includegraphics[width=0.49\textwidth,height=.45\textwidth]{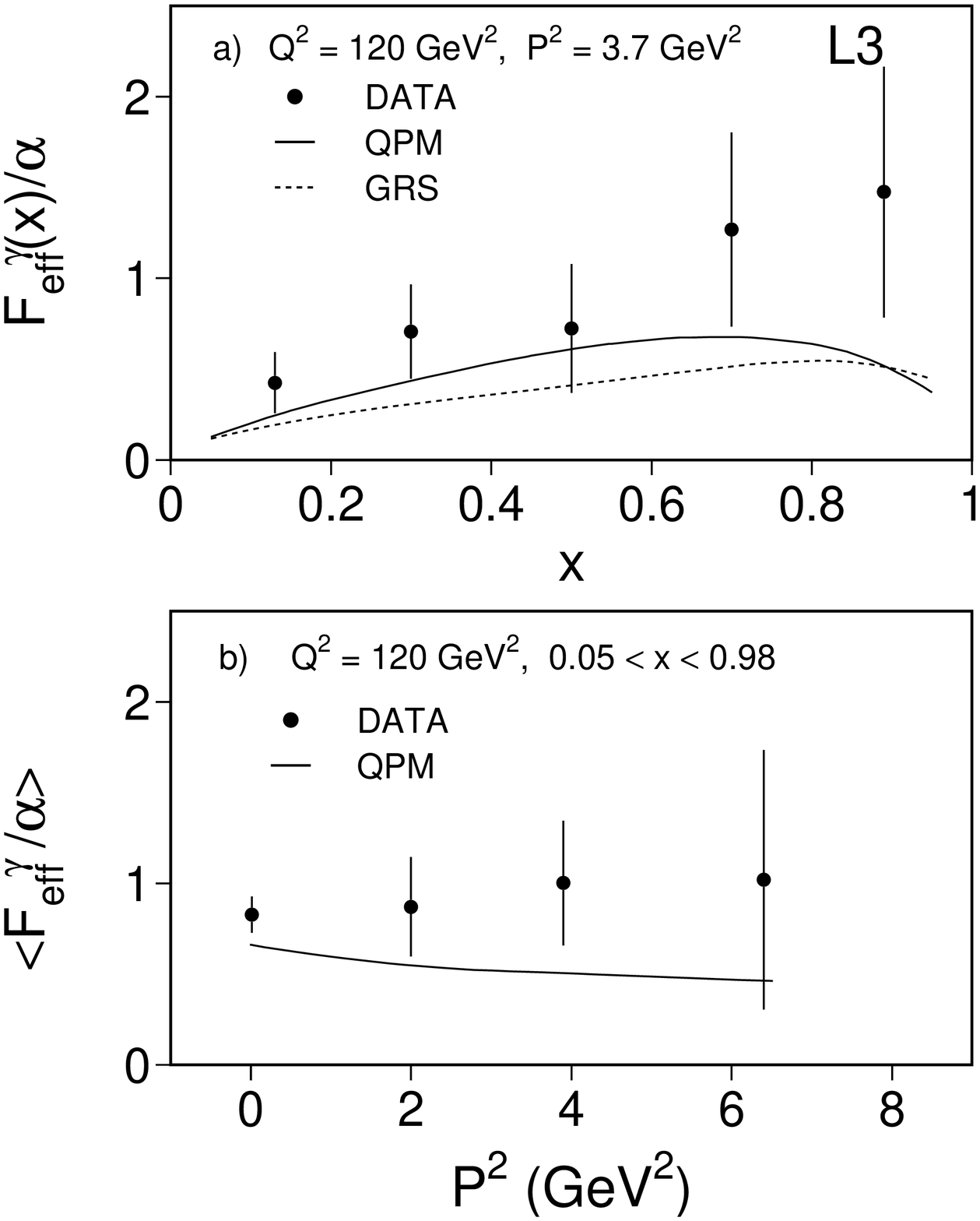}\vspace*{1mm}
\includegraphics[width=0.49\textwidth,height=.47\textwidth]{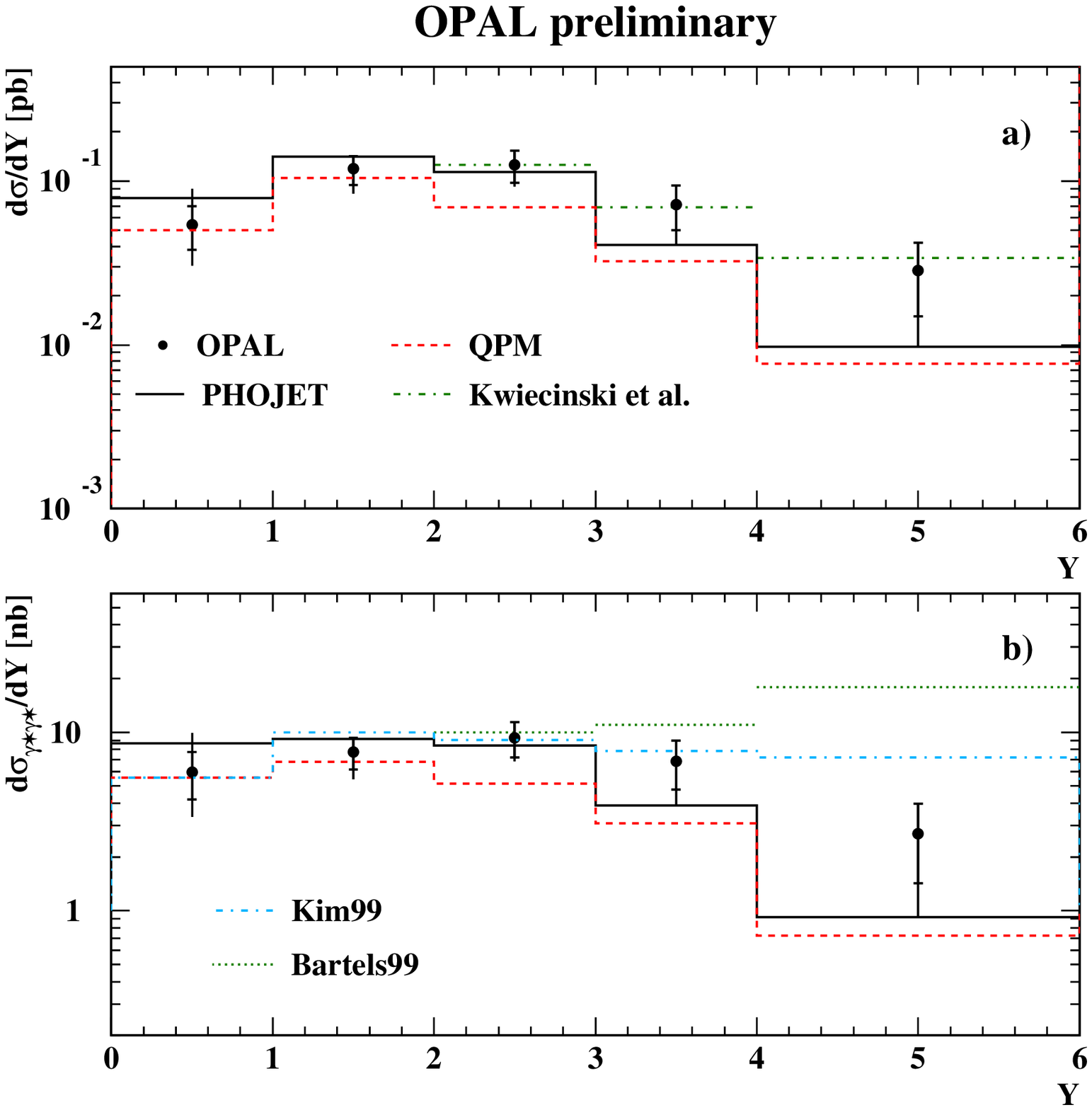}
\fcaption{The effective structure of the virtual photon measured by L3 (left)
and the double-tag cross-section as a function of  $Y\simeq
\ln{(W_{\gamma\gamma}^2 / \sqrt{Q^2_1Q^2_2})}$  measured by OPAL (right). }
\label{fig:doubletag}
\end{figure}

The charm structure function of the photon has also been
measured\cite{opaltagdstar} by OPAL using the D$^*$-tag method mentioned above
to identify charm production in single-tag DIS events. This first measurement
is well described by the theoretical expectations for $x>0.1$, while at lower
$x$ values the predictions are  below the data, with large statistical and
extrapolation uncertainties.

L3 has measured the double-tag cross-section in terms of the effective
structure function\cite{L3highQ2}, shown in  Figure~\ref{fig:doubletag} (left),
and in terms of the double-tag cross-section of two highly virtual
photons\cite{L3doubletag}, similar to the OPAL measurement\cite{opaldoubletag}
presented in Figure~\ref{fig:doubletag} (right). In all these measurements the
quark parton model calculation lies below the data, indicating contributions
from additional processes, while the LO BFKL prediction overestimates the data,
indicating important higher order corrections.

\section{Conclusions}
\noindent 
Several new and improved measurements of photon-initiated  QCD processes at LEP
have come out recently. In most cases we find 
a reasonable agreement with leading-order
Monte Carlo models, 
and a good agreement with NLO QCD calculations. The vast
amount of data collected by LEP promises more interesting results to come.

\nonumsection{References}

\end{document}